\documentclass[amsmath,amssymb,showkeys,superscriptaddress,floatfix,prd,10pt,aps]{revtex4-2}
\usepackage{graphicx,epstopdf}
\usepackage[caption=false]{subfig}
\usepackage{multirow}
\usepackage{appendix}
\def\xslash{x\!\!\!\slash }

\def\vel{\left|}
\def\ver{\right|}
\usepackage{colortbl}
\usepackage{xcolor}
\usepackage[colorlinks=true, urlcolor=blue, linkcolor=blue, citecolor=green]{hyperref}

\begin{document}

\title{Electromagnetic properties of the $ D \bar D^* K$  molecular hexaquark state}
\author{Ula\c{s} \"{O}zdem}%
\email[]{ulasozdem@aydin.edu.tr}
\affiliation{ Health Services Vocational School of Higher Education, Istanbul Aydin University, Sefakoy-Kucukcekmece, 34295 Istanbul, Turkey}

 
\begin{abstract}
We systematically study the electromagnetic properties of multiquark states. In this study, inspired by the recent series of studies that
showed the likely existence of a $ D \bar D^* K$ state, we examine the magnetic moment of  $ D \bar D^* K$  hexaquark state in three-meson molecular structure, as well as having isospin and spin-parity quantum numbers $I(J^P) =3/2(1^-)$   via light-cone sum rules. The magnetic moment obtained for the $ D \bar D^* K$ molecular hexaquark state is quite large due to the double electric charge, and its magnitude indicates that it is accessible in future experiments.  As a byproduct, the quadrupole  moment of the $ D \bar D^* K$ molecular hexaquark state is also extracted. This value indicates a non-spherical charge distribution.   The magnetic moments of hadrons contains valuable knowledge on the distributions of charge and magnetization their inside, which can be used to better understand their geometric shape and quark-gluon organizations. The results given in this study constitute an estimate of the magnetic moment of this $D \bar D^* K$ state and should serve as an inspiration to conduct experimental examinations of this state.
\end{abstract}
\keywords{Magnetic moment, $ D \bar D^* K$ molecular hexaquark state, Light-cone sum rules}

\maketitle

\section{Introduction}
Since the discovery of the $X(3872)$ by Belle in 2003~\cite{Belle:2003nnu}, there have been numerous candidates of multiquark states observed in particle experiments, which cannot be well explained in the conventional quark model. A lot of experimental and theoretical research has been done in the last few decades, but their nature is still elusive. We refer
to the reviews~\cite{Faccini:2012pj,Esposito:2014rxa,Chen:2016qju,Ali:2017jda,Esposito:2016noz,Olsen:2017bmm,Lebed:2016hpi,Guo:2017jvc,Nielsen:2009uh,Brambilla:2019esw,Liu:2019zoy, Agaev:2020zad, Dong:2021juy} and references therein for detailed discussions.

Theoretically, the attraction in the $DK$ and $D \bar D^*$ subsystems constitutes an argument in favor of the existence of $ D \bar D^* K$ state. The observation of such $ D \bar D^* K$ state might be possible in the current experimental facilities and it would constitute an exciting improvement in the multiquark states. This state, if found experimentally, definitely cannot be accommodated in a conventional meson picture, and hence presents a clear case of an multiquark state. In Ref. \cite{Ma:2017ery}, based on the attractive force of the
isosinglet $D^*K$ system the mass of the $ D \bar D^* K$ system has been studied using Born-Oppenheimer approximation via delocalized $\pi$ bond with isospin and spin-parity quantum numbers $I(J^P)= 1/2(1^-)$. 
In Ref. \cite{Ren:2018pcd}, the authors solved the Faddev equation to obtain the mass and decay width of $ D \bar D^* K$ state. The obtained mass and decay width are given as: $m = (4307 \pm 2) $ MeV and $\Gamma = (9 \pm 2)$ MeV with $I(J^P)= 1/2(1^-)$, respectively.
 In Ref. \cite{Ren:2019umd}, the possible two-body decay channels of $ D \bar D^* K$ state have been investigated via triangle diagrams with $I(J^P)= 1/2(1^-)$.
 In Ref. \cite{Di:2019jsx}, the spectroscopic parameters of $ D \bar D^* K$ system have been obtained in the framework of the QCD sum rules with quantum numbers $I(J^P)= 3/2(1^-)$. 
 In Ref. \cite{Wu:2020job}, the mass of the $ D \bar D^* K$ state have been investigated by solving the Schrödinger equation with quantum numbers $I(J^P)= 1/2(1^-)$.

There are different approaches in investigation of the structure of multiquark states, especially, the most promising one in this direction is the study of the electromagnetic properties of the multiquark states. Electromagnetic properties of the multiquark states play a vital role in understanding their internal structure and shape deformations. Besides, examining the magnetic dipole and higher multipole moments of multiquark states is a crucial element of understanding the heavy quark dynamics; however, the studies on the magnetic moments of multiquark states are scarce. 
Motivated by these facts, the magnetic moment of the $ D \bar D^* K$ vector hexaquark state has been obtained by using the light-cone sum rules~\cite{Chernyak:1990ag, Braun:1988qv, Balitsky:1989ry}. While obtaining this magnetic moment, it is taken into account that these state is in the three-meson molecular structure, as well as having $I(J^P) =3/2(1^-)$ quantum numbers.
Light-cone sum rules have been proven to be a powerful and successful nonperturbative method over the past few decades. Its starting point is to construct the corresponding interpolating currents appropriate to the hadron of concern, which have the prominent knowledge about the related hadron, like quantum numbers and the constituent particles.  
With these interpolating currents, the correlation function in the presence of the external electromagnetic field, which has two representations, the QCD representation and the hadronic representation, can be constructed. Matching these two representations via quark-hadron duality, the light-cone sum rules will be established, from which the magnetic moments, form factors, etc. may be deduced.

The rest of article is arranged as follows: After the introduction, in Sect. II we derive the light-cone sum rules for the magnetic moment of the $ D \bar D^* K$  hexaquark molecular state. In Sect. III, we present the numerical results and discussions. The summary of the paper is given in the Sect. IV.
 
 \begin{widetext}

 \section{Formalism }

The starting point of the light-cone sum rules is the correlation function constructed from two hadronic currents in the presence of the external electromagnetic field with the following form:
\begin{equation}
 \label{edmn01}
\Pi _{\mu \nu }(p,q)=i\int d^{4}xe^{ip\cdot x}\langle 0|\mathcal{T}\{J_{\mu}(x)
J_{\nu }^{ \dagger }(0)\}|0\rangle_{\gamma}, 
\end{equation}%
where $\mathcal{T}$, $J_{\mu}(x)$ and $\gamma$ represent the time-ordered product of two currents, the interpolating current of $ D \bar D^* K$ state and the external electromagnetic field, respectively. We need explicit expressions for $J_{\mu}(x)$ to make progress in the calculations. In the three-meson state, $J_{\mu}(x)$ with quantum numbers $I(J^P)= 3/2(1^-)$ can be written in the following form
\begin{eqnarray}\label{current}
J_\mu(x)&=& \Big\{\Big[\bar{u}^a(x)i\gamma_5 c^a(x) \Big] \Big[\bar{c}^b(x)\gamma_\mu d^b(x)\Big] \Big[\bar{u}^c(x)i\gamma_5 s^c(x) \Big] \Big\},
\end{eqnarray}
where the  $a$, $b$, $c$ denote color indices and the C is the charge conjugation matrix.

As the hadronic representation of the correlation function is concerned, we insert a complete set of intermediate hadronic state with same quantum numbers as the interpolating currents into the correlation function and we acquire
\begin{align}
\label{edmn04}
\Pi_{\mu\nu}^{Had} (p,q) &= {\frac{\langle 0 \mid J_\mu (x) \mid
D \bar D^* K(p, \varepsilon^\theta) \rangle}{p^2 - m_{D \bar D^* K}^2}} \langle D \bar D^* K(p, \varepsilon^\theta) \mid D \bar D^* K(p+q, \varepsilon^\delta) \rangle_\gamma
\frac{\langle D \bar D^* K(p+q,\varepsilon^\delta) \mid {J_\nu^{ \dagger}} (0) \mid 0 \rangle}{(p+q)^2 - m_{D \bar D^* K}^2} \nonumber\\
&+ \mbox{higher states}.
\end{align}
 
The amplitude $\langle 0 \mid J_\mu(x) \mid D \bar D^* K(p,\varepsilon^\theta) \rangle$ can be parameterized in terms of residue $\lambda_{D \bar D^* K}$ and polarization vector $\varepsilon_\mu^{\theta}$  of  $ D \bar D^* K$ state as
\begin{align}
\label{edmn05}
\langle 0 \mid J_\mu(x) \mid D \bar D^* K(p,\varepsilon^\theta) \rangle &= \lambda_{D \bar D^* K} \varepsilon_\mu^\theta\,,
\end{align}
while  the matrix element $\langle D \bar D^* K(p,\varepsilon^\theta) \mid  D \bar D^* K (p+q,\varepsilon^{\delta})\rangle_\gamma$ is given by 
\begin{align}
\langle D \bar D^* K(p,\varepsilon^\theta) \mid  D \bar D^* K (p+q,\varepsilon^{\delta})\rangle_\gamma &= - \varepsilon^\tau (\varepsilon^{\theta})^\alpha (\varepsilon^{\delta})^\beta \bigg\{ G_1(Q^2)~ (2p+q)_\tau ~g_{\alpha\beta}  + G_2(Q^2)~ ( g_{\tau\beta}~ q_\alpha -  g_{\tau\alpha}~ q_\beta) \nonumber\\ &- \frac{1}{2 m_{D \bar D^* K}^2} G_3(Q^2)~ (2p+q)_\tau ~q_\alpha q_\beta  \bigg\},\label{edmn06}
\end{align}
where $\varepsilon^\tau$ is polarization of the photon  and   $G_i(Q^2)$'s are electromagnetic form factors,  with  $Q^2=-q^2$.

Using Eqs. (\ref{edmn04})-(\ref{edmn06}) and after doing some necessary calculations the final form of the correlation function is  obtained as
\begin{align}
\label{edmn09}
 \Pi_{\mu\nu}^{Had}(p,q) &=  \frac{\varepsilon_\rho \, \lambda_{D \bar D^* K}^2}{ [m_{D \bar D^* K}^2 - (p+q)^2][m_{D \bar D^* K}^2 - p^2]}
 \bigg\{G_1(Q^2)(2p+q)_\rho\bigg(g_{\mu\nu}-\frac{p_\mu p_\nu}{m_{D \bar D^* K}^2}
 -\frac{(p+q)_\mu (p+q)_\nu}{m_{D \bar D^* K}^2}\nonumber\\
 &+\frac{(p+q)_\mu p_\nu}{2m_{D \bar D^* K}^4} (Q^2+2m_{D \bar D^* K}^2)
 \bigg)
 + G_2 (Q^2) \bigg(q_\mu g_{\rho\nu} - q_\nu g_{\rho\mu} -
\frac{p_\nu}{m_{D \bar D^* K}^2}  \big(q_\mu p_\rho - \frac{1}{2}
Q^2 g_{\mu\rho}\big) 
\nonumber\\
&+
\frac{(p+q)_\mu}{m_{D \bar D^* K}^2}  \big(q_\nu (p+q)_\rho+ \frac{1}{2}
Q^2 g_{\nu\rho}\big) -  
\frac{(p+q)_\mu p_\nu p_\rho}{m_{D \bar D^* K}^4} \, Q^2
\bigg)\nonumber\\
&
-\frac{G_3(Q^2)}{m_{D \bar D^* K}^2}(2p+q)_\rho \bigg(
q_\mu q_\nu -\frac{p_\mu q_\nu}{2 m_{D \bar D^* K}^2} Q^2 +\frac{(p+q)_\mu q_\nu}{2 m_{D \bar D^* K}^2} Q^2
-\frac{(p+q)_\mu q_\nu}{4 m_{D \bar D^* K}^4} Q^4\bigg)
\bigg\}\,.
\end{align}

To calculate the magnetic moment, we need only $G_2(Q^2)$ of the form factors described above. 
The magnetic form factor, $F_M(Q^2)$, is written as follows
\begin{align}
\label{edmn07}
&F_M(Q^2) = G_2(Q^2)\,.
\end{align} 

 The $F_M(Q^2=0)$ is proportional to the magnetic moment $\mu_{D \bar D^* K}$:
\begin{align}
\label{edmn08}
&\mu_{D \bar D^* K} = \frac{ e}{2\, m_{D \bar D^* K}} \,F_M(0).
\end{align}

To evaluate the QCD representation of the light-cone sum rules, we insert the interpolating currents to the correlation function, contract relevant quark fields and obtain QCD representation of the correlation function in terms of the light and heavy quark propagators
\begin{eqnarray} \label{QCDside}
\Pi^{QCD}_{\mu \nu}(p,q)&=&-i\int d^{4}x e^{ip\cdot x} \langle 0 \mid 
\Bigg\{{\rm Tr}[\gamma_\mu S_d^{bb'}(x)\gamma_\nu S_c^{b'b}(-x)]
{\rm Tr}[\gamma_5 S_c^{aa'}(x)\gamma_5 S_u^{a'a}(-x)]
\nonumber\\
&& \times {\rm Tr}[\gamma_5S_s^{cc'}(x)\gamma_5S_u^{c'c}(-x)] -{\rm Tr}[\gamma_\mu S_d^{bb'}(x)\gamma_\nu S_c^{b'b}(-x)]
{\rm Tr}[\gamma_5 S_c^{aa'}(x)i\gamma_5  
\nonumber\\
&& \times
S_u^{a'c}(-x)
\gamma_5 S_s^{cc'}(x)i\gamma_5 S_u^{c'a}(-x)]
\Bigg\} \mid 0 \rangle _\gamma\,,
\end{eqnarray}
 where 
$S_{q}(x)$ and $S_{c}(x)$ are represent the full light and massive quark propagators.  
Throughout our calculations, we use the x-space expressions for the $S_{q}(x)$ and $S_{c}(x)$~\cite{Yang:1993bp, Belyaev:1985wza}:
\begin{align}
\label{edmn12}
S_{q}(x)&=i \frac{{\xslash}}{2\pi ^{2}x^{4}} 
- \frac{\langle \bar qq \rangle }{12} \Big(1-i\frac{m_{q} \xslash}{4}   \Big)
- \frac{ \langle \bar qq \rangle }{192}m_0^2 x^2 
\Big(1 -i\frac{m_{q} \xslash}{6}   \Big)
-\frac {i g_s }{32 \pi^2 x^2} ~G^{\mu \nu} (x) \Big[\rlap/{x} 
\sigma_{\mu \nu} +  \sigma_{\mu \nu} \rlap/{x}
 \Big],
\end{align}
\begin{align}
\label{edmn13}
S_{c}(x)&=\frac{m_{c}^{2}}{4 \pi^{2}} \Bigg[ \frac{K_{1}\Big(m_{c}\sqrt{-x^{2}}\Big) }{\sqrt{-x^{2}}}
+i\frac{{\xslash}~K_{2}\Big( m_{c}\sqrt{-x^{2}}\Big)}
{(\sqrt{-x^{2}})^{2}}\Bigg]
-\frac{g_{s}m_{c}}{16\pi ^{2}} \int_0^1 dv\, G^{\mu \nu }(vx)\Bigg[ \big(\sigma _{\mu \nu }{\xslash}
  +{\xslash}\sigma _{\mu \nu }\big)\nonumber\\
  &\times \frac{K_{1}\Big( m_{c}\sqrt{-x^{2}}\Big) }{\sqrt{-x^{2}}}
+2\sigma_{\mu \nu }K_{0}\Big( m_{c}\sqrt{-x^{2}}\Big)\Bigg].
\end{align}%
where $\langle \bar qq \rangle$ is quark  condensate, $m_0$ is defined through the quark-gluon mixed condensate  $\langle 0 \mid \bar  q\, g_s\, \sigma_{\alpha\beta}\, G^{\alpha\beta}\, q \mid 0 \rangle = m_0^2 \,\langle \bar qq \rangle $, $v$ is line variable, $G^{\mu\nu}$ is the gluon field strength tensor,  $\sigma_{\mu\nu}= \frac{i}{2}[\gamma_\mu, \gamma_\nu]$ and $K_i$'s are modified Bessel functions of the second kind. 
The first term of the light and heavy quark propagators correspond to perturbative or free part and the rest belong to the interacting parts. 

A few remarks should be made here regarding the calculation of the QCD representation of the correlation function. 
The correlation function in Eq. (\ref{QCDside}) includes two different contributions as perturbative and nonperturbative.  
In practice, perturbative contributions, the photon interacts with one of the quarks, can be computed by the replacing the one of the light or heavy-quark propagators by
\begin{align}\label{lightprop}
S^{free} \rightarrow \int d^4z\, S^{free} (x-z)\,\rlap/{\!A}(z)\, S^{free} (z)\,.
\end{align}
The remaining five propagators are considered as full propagators. 
Nonperturbative contributions, the photon is radiated at long distances, can be computed by replacing one of the light quark propagators by 
\begin{align}\label{cquarkprop}
S_{\alpha\beta}^{ab} \rightarrow -\frac{1}{4} (\bar{q}^a \Gamma_i q^b)(\Gamma_i)_{\alpha\beta},
\end{align}
where $\Gamma_i$ are full set of the Dirac matrices and surviving  propagators are considered as full quark propagators.
It is seen that matrix elements such as $\langle \gamma(q) | \bar{q}(x) \Gamma_i q(0) | 0\rangle$ and $\langle \gamma(q)\vel \bar{q}(x) \Gamma_i G_{\mu\nu}q(0) \ver 0\rangle$ emerge after the calculations given in Eq. (\ref{cquarkprop}) are made. 
These matrix elements are parameterized with respect to the photon distribution amplitudes (DAs), which are
the key nonperturbative parameters in light-cone sum rules, whose explicit expressions are
given 
in Ref.~\cite{Ball:2002ps}. Besides these matrix elements, in principle, nonlocal operators such as $\bar q G^2 q$ and $ \bar q q \bar q q$ are expected to show up. However, it is known that the contributions of such operators are small, which is supported by the conformal spin expansion \cite{Balitsky:1987bk, Braun:1989iv}, and hence we will omit them.
When the above-mentioned calculations are made, the QCD representation of the correlation function is obtained.

Finally, the structure $q_\mu \varepsilon_\nu$ is chosen from both representations and the coefficients of the structure are matched in both hadronic and QCD representations. Then, Borel transformation and continuum subtraction are used to suppress the effects of the continuum and higher states. As a result, the light-cone sum rule for $ D \bar D^* K$ state is as follows:
\begin{align} \label{sonn}
 &\mu_{D \bar D^* K} \, \lambda_{D \bar D^* K}^2  = e^{\frac{m_{D \bar D^* K}^2}{M^2}} \,\, \Delta^{QCD} (M^2, s_0),
\end{align}
where $M^2$ and $s_0$ are free parameters originating from the applications of the Borel transformation and continuum subtraction procedures. 
The $\Delta^{QCD}(M^2,s_0)$  function is rather lengthy,  explicit expression of which is not given in the text. 
%
In obtaining the final result, we employ  $ \frac {1}{M^2} =\frac {1}{M_1^2}+\frac {1}{M_2^2}$ with $ M_1^2 $ and $ M_2^2 $ being the Borel parameters in the initial and final states, respectively. We set $ M_1^2= M_2^2 =2 M^2 $  as the initial and final states are the same. We will fix $ M^2 $ and continuum threshold ($ s_0 $)  based on the standard prescription of the light-cone sum rule method in next section.

At the end of this section, we would like to note that the magnetic moment of the $ D \bar D^* K$ state has been computed from the light-cone sum rules employing for its hadronic side a single-pole approach [see, Eq. (\ref{edmn04})]. In the case of the multiquark hadrons such approach should be verified by additional arguments, because a hadronic side of relevant sum rules receives contributions from two-hadron reducible terms as well. Two-hadron contaminating terms have to be considered when extracting parameters of multiquark hadrons ~\cite{Kondo:2004cr,Lee:2004xk}. In the case of the multiquark hadrons they lead to modification in the quark propagator
\begin{equation}
\frac{1}{m^{2}-p^{2}}\rightarrow \frac{1}{m^{2}-p^{2}-i\sqrt{p^{2}}\Gamma (p)%
},  \label{eq:Modif}
\end{equation}%
where $\Gamma (p)$ is the finite width of the  multiquark hadrons generated by two-hadron scattering states.  When these contributions are properly taken into account in the sum rules,  they rescale the residue of the  multiquark hadrons under investigations leaving its mass unchanged. 
Detailed investigations show that two-hadron scattering contributions are small for  multiquark hadrons (see Refs. \cite{Wang:2015nwa,Agaev:2018vag,Sundu:2018nxt,Wang:2019hyc,Albuquerque:2021tqd,Albuquerque:2020hio,Wang:2020iqt,Wang:2019igl,Wang:2020cme}).
Therefore, one can safely exclude the contributions of two-hadron scattering effects in the hadronic side of the correlation function.

\end{widetext}

\section{Numerical analysis and conclusions}

We assume the following parameters to perform the numerical calculations for the magnetic moment of the $D \bar D^* K$ state. The masses of the light quarks are $m_u=m_d=0$, $m_s =96^{+8}_{-4}\,\mbox{MeV}$,  the mass of the c-quark is $m_c = (1.275\pm 0.025)\,$GeV, the condensates of the light quarks are  $\langle \bar ss\rangle = 0.8\, \langle \bar uu\rangle$  with $\langle \bar uu\rangle $
=$\langle \bar dd \rangle $=$(-0.24\pm0.01)^3\,$GeV$^3$~\cite{Ioffe:2005ym}, the gluon condensate is $\langle g_s^2G^2\rangle = 0.88~ \mbox{GeV}^4$~\cite{Matheus:2006xi} and the quark-gluon condensate is $m_0^{2} = 0.8 \pm 0.1$~GeV$^2$~\cite{Ioffe:2005ym}. 
To proceed the numerical calculations of the magnetic moment of the $D \bar D^* K $ state , numerical values of the mass and residue parameters of this state are also needed. These values have been computed in Ref.~\cite{Di:2019jsx}.  The obtained results for mass and residue are given as $m_{D \bar D^* K} = 4.71^{+0.19}_{-0.11}~ \mbox{GeV}$ and $\lambda_{D \bar D^* K} = (4.60^{+1.15}_{-0.69}) \times 10^{-4}~ \mbox{GeV}^8$.
All necessary terms regarding the DAs of the photon are borrowed from Ref.~\cite{Ball:2002ps}.

The light-cone sum rules in Eq. (\ref{sonn}) is a function of the Borel mass parameter $M^2$ and continuum threshold $s_0$. To acquire a reliable light-cone sum rules result, one should  determine proper working intervals for these two free parameters. The OPE convergence and pole dominance (PC) constraints are widely used to determine the working intervals of these two parameters. Considering these limitations, the following working  intervals are obtained for these two free parameters as a result of the numerical analysis,
\begin{eqnarray}
 5.0~\mbox{GeV}^2 \leq M^2 \leq 7.0~\mbox{GeV}^2 \nonumber\\
  25.2~\mbox{GeV}^2 \leq s_0 \leq 27.2~\mbox{GeV}^2. 
\end{eqnarray}

Using the above working intervals for the $M^2$ and $s_0$, the PC varies in the intervals $53 \% \geq$ PC $\geq 15 \%$. At $M^2_{max} =7.0~$GeV$^2$, the PC is equal to $15\%$, while at $M^2_{min} =5.0~$GeV$^2$, it is equal to $53\%$. In the standard analysis of QCD sum rules, the PC should be larger than $50\%$ for baryons and mesons. In the case of tetra- and pentaquark states, it turns out to be as PC $ > 20\%$. In Refs. \cite{Mutuk:2022zgn,Chen:2021hxs, Chen:2019vdh, Chen:2014vha}, it is indicated that hexaquark spectral densities led to small PC.   
When we investigate the OPE convergence, we have acquired that the contribution of the higher dimensional terms in OPE is less than $\sim 2 \%$,  thus the convergence of the sum rules is ensured.  
In Fig. 1, the $M^2$ and $s_0$ dependencies of the magnetic moment of the $ D \bar D^* K$ hexaquark state is shown.   As is seen, the variation of the results with respect to the continuum threshold is substantial; however, there is much less dependence of the magnetic moment on the Borel parameters in its working interval.

  \begin{figure}[htp]
  \label{aafig}
\centering
 \includegraphics[width=0.45\textwidth]{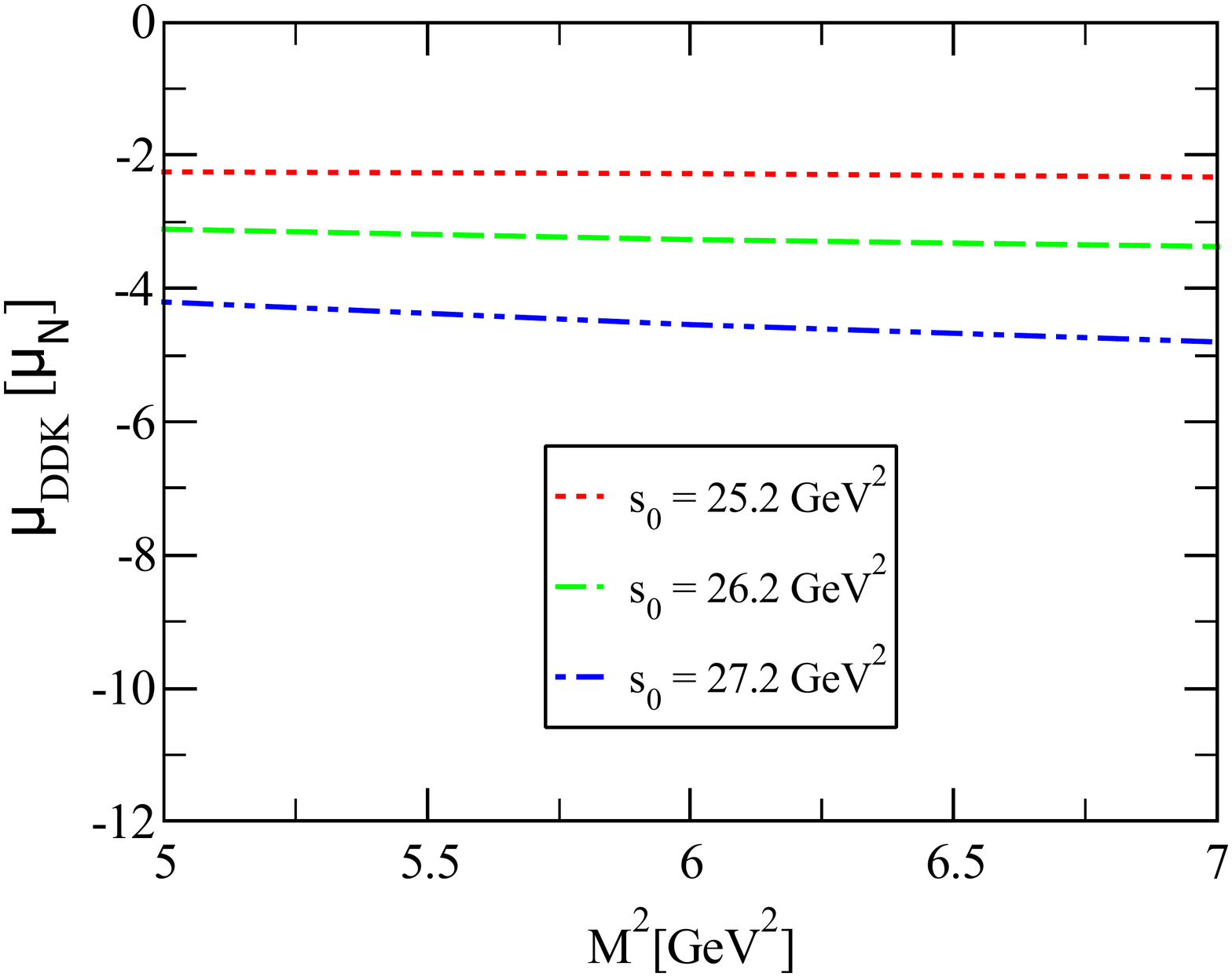}
 \includegraphics[width=0.45\textwidth]{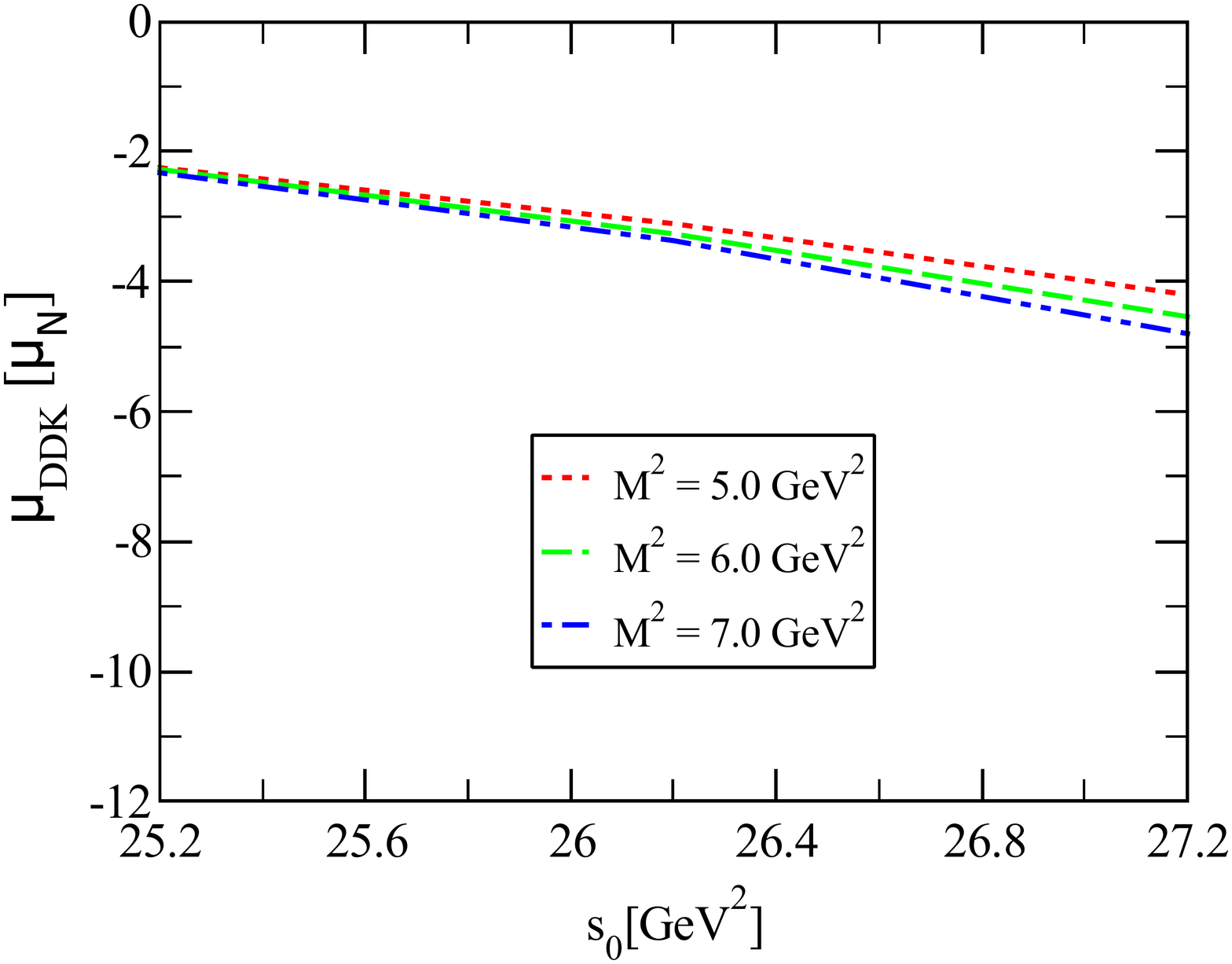}
 \caption{ 
 Variations of the magnetic moment of $ D \bar D^* K$ state with $M^2$ and $s_0$.}
  \end{figure}

After all the above procedures are completed, the results we obtained for the magnetic moment are given as follows

\begin{align} \label{mmom}
 \mu_{ D \bar D^* K} &= - 3.52 \pm 1.28 \, \mu_N.
\end{align}
The error in the Eq.~(\ref{mmom}) is due to all input parameters, extra parameters such as $s_0$ and $M^2$, as well as numerical parameters used in expressions in photon DAs. We see from this result that, the value of the magnetic moment of $ D \bar D^* K$ state are quite large because of the double electric charge.

As a byproduct, we have also acquired  quadrupole moment ($\mathcal{D}$) of the $D \bar D^* K$ state as follows
\begin{align} \label{Qmom}
  \mathcal{D}_{ D \bar D^* K} &= - 0.030 \pm 0.008 \,fm^2.
\end{align}
This result indicates that the charge distribution of the $D \bar D^* K$ state is non-spherical. 

A comparison of our predictions on the magnetic and quadrupole moments of the $ D \bar D^* K$ state with the estimations of other methods, such as lattice QCD, quark model, chiral perturbation theory and so on  would be interesting. As we mentioned above $D \bar D^* K$ state belong to a class of doubly charged multiquark states that the measurements of their electromagnetic parameters, like those of the $\Delta^{++}$ baryon, are relatively easy compared to other multiquark states. These kind of multiquark states have not been explored so far. We hope our estimations on the electromagnetic properties of these states together with the results of other theoretical studies on the spectroscopic parameters of these states will be useful for their searches in future experiments and will help us define exact inner structures of these multiquark states.

Let us briefly discuss how the magnetic moment of this state can be measured experimentally. The possible short lifetime of $ D \bar D^* K$ state does not allow the employ of  spin procession technique to measure the magnetic and quadrupole moments.  An alternative technique based on photon emission off hadrons \cite{Zakharov:1968fb} can be used in the later case, since the photon carries data on higher multipole moments of emitting hadrons. The basic scheme of this idea is that the amplitude for radiative process can be expressed as a power expansion in the photon energy $\omega_\gamma$ as follows
\begin{align}
 M \sim A\,(\omega_\gamma )^{-1} + B\,(\omega_\gamma )^0 +\cdots,
\end{align}
where $(\omega_\gamma )^{-1}$, $(\omega_\gamma)^0$  and dots represent the electric charge, magnetic moment and higher multipole moments, respectively.
By measuring the cross section or decay width of the radiative process and excluding from the small contributions of terms linear/higher order in $\omega_\gamma$, one can define the magnetic moment of related hadron.

 \section{Summary} \label{summ}
 
 We systematically study the electromagnetic properties of multiquark states. In this study, inspired by the recent series of studies that
showed the likely existence of a $ D \bar D^* K$ state, we examine the magnetic moment of  $ D \bar D^* K$  hexaquark state in three-meson molecular structure, as well as having isospin and spin-parity quantum numbers $I(J^P) =3/2(1^-)$   via light-cone sum rules.
The magnetic moment obtained for the $ D \bar D^* K$ molecular hexaquark state is quite large due to the double electric charge, and its magnitude indicates that it is accessible in future experiments.  As a byproduct, the quadrupole  moment of the $ D \bar D^* K$ molecular hexaquark state are also extracted. This value indicates a non-spherical charge distribution.   The magnetic moments of hadrons contain valuable knowledge on the distributions of charge and magnetization their inside, which can be used to better understand their geometric shape and quark-gluon organizations. The results given in this study constitute an estimate of the magnetic moment of this $D \bar D^* K$ state and should serve as an inspiration to conduct experimental examinations of this state.

\textbf{Data Availability Statement:} This manuscript has no associated data or
the data will not be deposited. [Authors’ comment: This is a theoretical
research work, so no additional data are associated with this work.]

\bibliography{DDbarK}

\end{document}